\documentclass[10pt, conference]{IEEEtran}

\usepackage{url}
\usepackage{todo}
\usepackage{color}
\usepackage{xspace}
\usepackage{amsthm}
\usepackage{hyperref}
\usepackage{multirow}
\usepackage{graphicx}
\usepackage{algorithm}
\usepackage{enumerate}
\usepackage{algpseudocode}
\usepackage{amsmath, amssymb, amsfonts}

\newcommand{\toolname}{LIRAT\xspace}
\newcommand{\image}{Widget Feature Matching}
\newcommand{\layout}{Layout Characterization Matching}

\newcommand{\tabincell}[2]{\begin{tabular}{@{}#1@{}}#2\end{tabular}}

\begin{document}

\title{Layout and Image Recognition Driving \\ Cross-Platform Automated Mobile Testing}

\author{Anonymous Author(s)}

 \author{\IEEEauthorblockN{Shengcheng Yu, Chunrong Fang$^*$, Yexiao Yun, Yang Feng}
 	\IEEEauthorblockA{State Key Laboratory for Novel Software Technology, Nanjing University, China \\
 	$^*$Corresponding author: fangchunrong@nju.edu.cn}}

\maketitle

\begin{abstract}

The fragmentation problem has extended from Android to different platforms, such as iOS, mobile web, and even mini-programs within some applications (app), like WeChat\footnote{A popular chatting app in China, providing a platform for other manufacturers to deploy mobile apps.}. In such a situation, recording and replaying test scripts is one of the most popular automated mobile app testing approaches. However, such approach encounters severe problems when crossing platforms. Different versions of the same app need to be developed to support different platforms relying on different platform supports. Therefore, mobile app developers need to develop and maintain test scripts for multiple platforms aimed at completely the same test requirements, greatly increasing testing costs. However, we discover that developers adopt highly similar user interface layouts for versions of the same app on different platforms. Such a phenomenon inspires us to replay test scripts from the perspective of similar UI layouts.

In this paper, we propose \textit{\textbf{an image-driven mobile app testing framework}}, utilizing \image~and \layout~to analyze app UIs. We use computer vision (CV) technologies to perform UI feature comparison and layout hierarchy extraction on mobile app screenshots to obtain UI structures containing rich contextual information of app widgets, including coordinates, relative relationship, etc. Based on acquired UI structures, we can form a platform-independent test script, and then locate the target widgets under test. Thus, the proposed framework non-intrusively replays test scripts according to a novel platform-independent test script model. We also design and implement a tool named \toolname to devote the proposed framework into practice, based on which, we conduct an empirical study to evaluate the effectiveness and usability of the proposed testing framework. The results show that the overall replay accuracy reaches around 65.85\% on Android (8.74\% improvement over state-of-the-art approaches) and 35.26\% on iOS (35\% improvement over state-of-the-art approaches).

\end{abstract}


\section{Introduction}

Fragmentation problem is proposed in \cite{wei2016taming}.
In the situation of Android fragmentation problems, recording and replaying test scripts on a large scale device cluster is one of the key quality assurance technologies for mobile apps. It can automatically execute preset test cases and detect various bugs \cite{qin2016mobiplay}. Test scenarios recorded on one device can be replayed on other devices of different hardware or software (e.g. operating system versions). Moreover, the fragmentation problem has extended to multiple platforms, including Android, iOS, mobile web, and even mini-programs within some apps, like WeChat. Here, we define the ``platform'' much more than operating system, but refer to a set of complete frameworks that independently provide an environment to support the apps to run. The feature of rapid iteration and frequent requirement change of mobile apps on different platforms triggers even increasing demands on app quality assurance. For a specific app, the expanded fragmentation problem on all mobile platforms means multiple clients for different platforms sharing the same services provided by a unified server. More importantly, they also share similar UI layouts for better human-computer interaction.

The expanded fragmentation problem raises a higher demand for developers when they test their mobile apps. In other words, they have to write different test scripts based on different framework supports for completely the same test requirements, leading to tedious and repetitive work. Moreover, different customized operating system versions can even have different supports for test script execution. This phenomenon causes a great burden on developers because they need to get familiar with platform-specific features. Besides, the test scripts are impossible to execute generally for different platforms. In lack of platform-free testing technologies, it is impossible for cross-platform test script record and replay.

\begin{figure}[!h]
	\centering
	\vspace{-0.1cm}
	\includegraphics[width=0.85\linewidth]{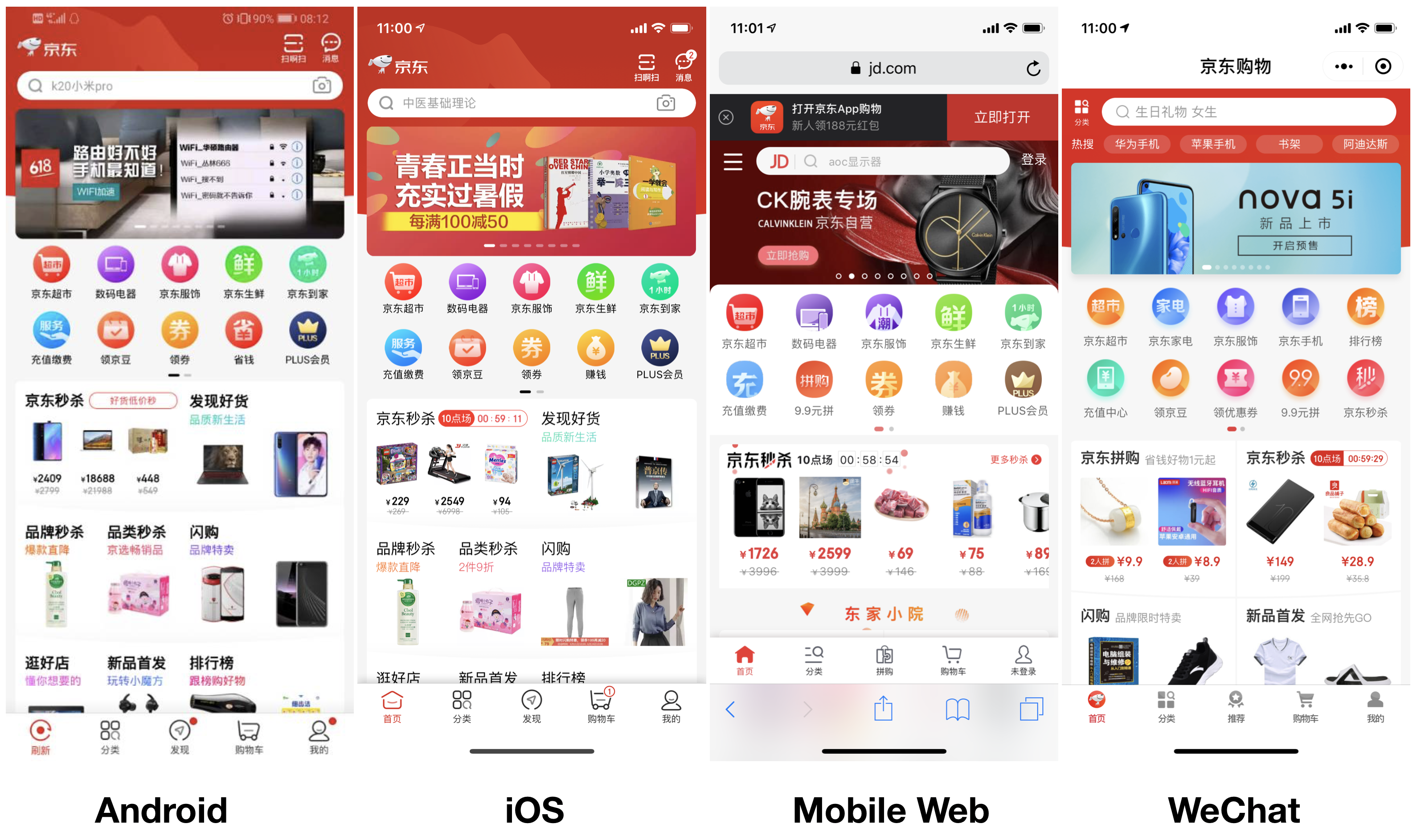}
	\caption{An Example: Multiple Clients for \textit{JingDong}}
	\label{fig:JD}
\end{figure}

Most mobile developers design the app user interface (UI) with a high similarity when supporting various platforms to improve the user experience. Here, we take one of the most popular online shopping apps in China (like Amazon in the US), \textit{JingDong}, as an example to illustrate such a phenomenon (See Fig. \ref{fig:JD}). It is evident from Fig. \ref{fig:JD} that the general layouts are with high similarity on different platforms, as are the UI elements and their relative positions, despite some slight differences on contents. However, faced with these highly similar UI layouts, developers are still required to develop test scripts respectively for different platforms based on platform-specific features to test the same functions, and to meet completely the same test requirements. They also need to adopt complete different testing frameworks on various platforms, which significantly increases both economic and human resource costs. The similar UI layouts enlighten us to research from the UI perspective instead of the respective underlying implementation.

To reduce the costs of adapting scripts among a wide range of platforms, some existing researches start with different points of views, like exploring the runtime statements, matching source code of the apps on various platforms, taking advantage of UI images, etc. However, such techniques tend to be intrusive and have much overhead \cite{singh2014comparative} \cite{guo2020crowdsourced}. Also, they can hardly identify dynamic or similar widgets, which are common in current mobile apps. Different platform features and different testing framework supports always make it a failure to relocate the target widgets and replay test scripts even on the same platform, let alone cross-platform replays.

Some researchers have done primary studies based on image understanding techniques. Sikuli \cite{sikuli} \cite{yeh2009sikuli} is an image-based testing tool focused on desktop apps, and it can identify and manipulate GUI widgets without source code. However, Sikuli has a poor support for the mobile environment. Sikuli relies on simple image feature matching, which can lead to failures when images are too simple to extract enough features to match. Airtest \cite{airtest} is another testing tool based on image-driven technologies, and it is developed on the basis of Sikuli. However, Airtest adopts different matching solutions for various mobile platforms, thus making it still hardly possible to record and replay test scripts among different platforms.

Tools mentioned above merely make simple image feature extraction and matching, making it tough to deal with dynamic elements, which is common in such a data explosion era. For example, in a news app, each piece of news has a different content, constructing a distinct image feature set. When the content is refreshed, traditional technologies will have trouble locating the recorded widgets via image features. Such tools take ``\texttt{images}'' as ``\texttt{images}'' only instead of ``\texttt{widget sets}''. In other words, they ignore the contents and mutual relationships, and some important information is left out.

In this paper, we propose an image-driven mobile app testing framework to solve the cross-platform record and replay problem of test scripts for the first time. We combine image context understanding and layout extraction to solve the problem. During the recording phase, the proposed testing framework automatically characterize the layout, and extracts widget screenshots, layout coordinates, and other attributes from the testing devices. With the obtained information, we form a test script according to the platform-independent test script model introduced in Section \ref{sec:test_script_model}, including all the extracted screenshots and well-organized hierarchy XML files. In the replaying phase, the proposed testing framework adopts both traditional computer vision and deep learning technologies. The image-driven mobile app testing framework is composed of \image~and \layout. \layout~can compensate for the drawbacks of \image~when a mobile app activity has dynamic or several similar widgets. After a comprehensive analysis of the intermediate results of \image~and \layout, \toolname can reach a high accuracy when positioning widgets on different devices. Therefore, the corresponding operations can be successfully replayed.

The proposed image-driven mobile app testing framework realizes ``one script record, multiple script replays'' on devices of different platforms. The framework utilizes the combination of image understanding and layout extraction for the first time, and the framework significantly reduces the complexity of test script developing. 

Based on the image-driven mobile app testing framework, we design and implement a tool named \toolname. \toolname simulates real app manipulation and simplifies the test script developing process into operation sequence record. Users operate on webpages, where a projection of real mobile devices are shown. Also, We conduct an empirical experiment to evaluate our image-driven mobile app testing framework on \toolname.

Main contributions of this paper are as follows:

\begin{itemize}
    \item We propose an image-driven mobile app testing framework for cross-platform test script record and replay, solving the problem of reusing test scripts cross platforms.
    \item We introduce a platform-independent test script model containing rich information recorded from mobile apps, including screenshots, widget information, etc.
    \item We declare a comprehensive cross-platform widget matching approach, including \image~and \layout, and based on which we design and implement a novel tool, which can record and replay test scripts on multiple platforms.
    \item We conduct an empirical experiment on how our approach with real-world apps and devices, and the tool can effectively improve testing efficiency.
\end{itemize}

The rest of this paper is organized as follows. Section \ref{sec:background&motivation} introduces the problems and existing solutions, together with their drawbacks. Section \ref{sec:methodology} illustrates the methodology in detail, including the pivotal technologies in the record phase and replay phase. The specific tool design and implementation are presented in Section \ref{sec:tool}. In Section \ref{sec:verification}, an empirical experiment is conducted to evaluate the effectiveness of the proposed approach. Section \ref{sec:relatedWork} introduces the related work. Finally, this research is concluded in Section \ref{sec:conclusion}.

More details of the framework design and the experiment data can be found on \url{https://sites.google.com/view/lit2020}.

\section{Background \& Motivation}
\label{sec:background&motivation}

\begin{figure*}[!h]
	\centering
	\includegraphics[width=0.9\linewidth]{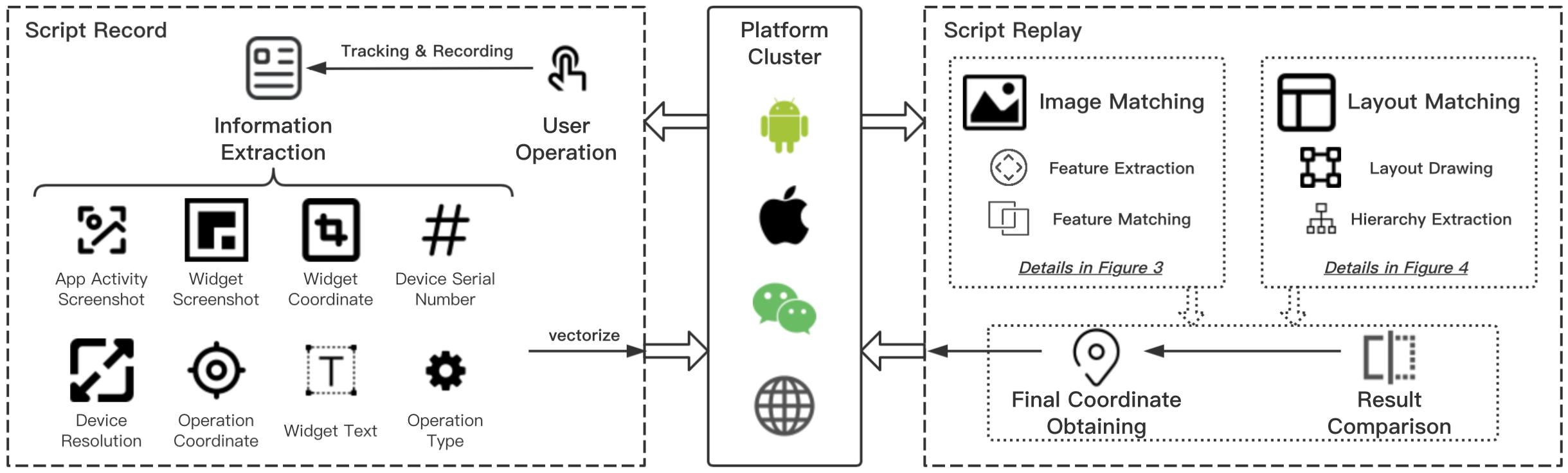}
	\caption{Image-Driven Mobile App Testing Framework}
	\label{fig:workflow}
\end{figure*} 

Many researches have been done to deal with the record and replay of test scripts. They start from different perspectives, including runtime statements, source codes, UI images, etc. However, such solutions still have drawbacks and limitations.

\subsection{Script Replay Dilemma}

The fragmentation problem originates from the Android platform, which is open for all manufacturers to make modifications to satisfy their own demands. The openness of Android has led to hundreds of thousands of different types of mobile models with different brands, operating system versions, screen shapes, resolutions, etc. Currently, the fragmentation problem has extended to many other platforms, such as 1) iOS, where devices have different screen shapes, different resolutions; 2) mini-programs within some apps, where mobile apps need to show the same appearances but base on different operating system kernels; and 3) mobile web, where mobile apps need to make special modifications to user interfaces for mobile web browsers. Developers have to develop different versions of an app to Adapt to different platforms.

We also conduct an analysis on the fragmentation problem. We find that 8 apps in the top 10 free apps and 7 apps in the top 10 paid apps in the Google Play \cite{googleplay} have their iOS versions, and most of them have mobile web versions\footnote{Some applications have no mobile web version due to their categories, such as tools, video games, etc.}. This result proves the fragmentation problem actually exists and has a deep impact on the mobile app developing.

Such extended fragmentation problem leads to a great burden for developers on app testing. Developing test scripts for each platform and each version respectively can be indeed painful and time-consumption, and makes it easy for developers to make mistakes. Each platform will require a group of developers to work on app testing. Moreover, as for developers, testing work is much more substantial. In traditional testing, developers must acquire the underlying information, such as the widget ID or XPath. This phenomenon makes it hard for cross-platform replay, because the implementations for multiple platforms of the same mobile app are quite different and based on the features of each platform.

\subsection{Limitations of Current Solutions}

Sikuli is an automated testing tool presented by Yeh et al. \cite{yeh2009sikuli}. Sikuli uses activity screenshots to generate test cases and execute automated testing for desktop apps. It can be used for various web-based apps and desktop applications \cite{singh2014comparative}. However, Sikuli has a poor support for mobile devices. Though it can be used to operate mobile device projections on desktop emulators, being unable able to operate on real devices is a significant drawback. Moreover, Sikuli adopt simple image feature extraction and matching, which is unsuitable for the constantly refreshing apps. Therefore, the problem of recording and replaying test scripts for mobile apps is still unresolved. However, the ideas of Sikuli inspired us to make use of mobile app UI elements.

Airtest is an automated testing tool for GUI testing. Airtest has a better support for mobile platforms. Airtest technology can acquire the whole UI tree structures from the \texttt{.apk} files, and identify the target widgets. Then, related simulative operations to replay scripts will be executed. Airtest mainly focuses on video game testing, where widgets have different image features. Therefore, Airtest still has problems when facing mobile apps of a wider range of categories, such as tools, news apps, etc., especially when the widgets have similar image features. Moreover, Airtest cannot execute cross-platform replays of the same script, such as iOS and mobile web client because it adopts different script developing implementations on different platforms.

\section{Methodology}
\label{sec:methodology}

Our proposed image-driven mobile app testing framework consists of two processes: \emph{Script Record} and \emph{Script Replay} (See Fig. \ref{fig:workflow}). Besides, the proposed framework also adopts a novel platform-independent test script model.

In the \emph{Script Record} process, the proposed framework records the screenshots and layout information of the widgets operated by the users required in the test requirements, and translates obtained information into test scripts according to the proposed platform-independent test script model $LITS$. 

In the \emph{Script Replay} process, we extracted the $LITS$ script, and combine \image~and \layout~to match the corresponding widgets on the replaying devices according to the screenshots and layout information recorded in the $LITS$ instances.

\subsection{Platform-Independent Test Script Model}
\label{sec:test_script_model}

We propose a novel test script model, named $LITS$, which means \textbf{\textit{Layout \& Image Test Script}}. $LITS$ is platform-independent because we extract and record all the information from UI screenshots without relying on platform-specific features or functions. Moreover, we make further processing to make the obtained information free of device attributes. Therefore, $LITS$ gets rid of both software and hardware dependence and can be used uniformly without adaptation.

During the script record, we extract the rich but necessary information after each operation. During the script replay, we also extract the same information from the replaying device and match with the information stored in $LITS$ scripts. The information for each operation includes 
the screenshot of the app activity under test, denoted as $SS_a$; 
the screenshot of the operated widget, denoted as $SS_w$; 
the coordinate of the widget, denoted as $C_w$, which is composed of the top-left and the bottom-right coordinates of the operated widget; 
the operated point coordinate, denoted as $C_o$; 
the recording device serial number, denoted as $DSN$; 
the recording device resolution, denoted as $DR$; 
texts on the widget, denoted as $T$; 
and the operation type, denoted as $O$. 

All the above information are critical to the cross-platform record and replay.
$SS_a$ is used to analyze the whole context of the activity, we can extract all the widgets, including $C_w$ from $SS_a$.
$SS_w$ is extracted from $SS_a$, representing the target widget. $SS_w$ plays an important role in \image~replay.
$C_w$ is representing with the distance from the top-left vertex of the screenshot, and it is used for the layout characterization.
$C_o$ is the operated point, and it can help judge the operated point falls in which extracted widget screenshot range.
$DSN$ is combined with the recording timestamp as the script id for store.
$DR$ is also an important element. It is used to calculate the relative proportional position of the widget and the operation point.
$T$ is the texts on the widget, which can assist identify the widgets.
$O$ refers to the operation type, like click, slide, etc., which is also an indispensable part.

When the above information is obtained, further processing is required for making the script platform-independent. First, we combine the $DSN$ and recording timestamp as the script id. Second, we calculate the relative proportional position of the widget and the operation point, specifically, which are calculated as the proportion of the absolute coordinate and the resolution of the device, and the results are denoted as $C_{wr}$, $C_{or}$. Therefore, $LITS$ is a list of 7-tuples, which is $ <ID, SS_a, SS_w, O, C_{wr}, C_{or}, T>$, and each 7-tuple represent for a user operation.

\subsection{Script Record}

Script record is implemented by a series of single-step operations record. The screenshots and layout information of the operated widgets are extracted and recorded for each operation as a $LITS$ instance, attached with some attribute information of the widgets, such as texts, widget types, etc.

When received by the recording device, the user's operations are converted into executable instructions on different platforms through the ADB \cite{adb} (for Android) or WDA \cite{wda} (for iOS). The extracted widget information, including coordinates, texts, etc. is recorded, and based on the information, XML files representing the activity layout in tree structures will be generated automatically through the proposed framework. Together with the activity screenshots and the widget screenshots cropped from activity screenshots, the XML files are stored in the form of a nested directory, the root directory represents for the test script, and each subdirectory represents for the widget information file for each operation, and the operation sequence is stored in the root directory. Algorithm \ref{alg:record} shows the formal expression. The input is a sequence of operations, denoted as $OS$, and the output is a platform-independent test script defined in Section \ref{sec:test_script_model}, denoted as $LITS$.

\textbf{\textit{Script Record}}: Based on $LITS$ model, each operation on a widget is recorded. Then, necessary information mentioned in Section \ref{sec:test_script_model} is automatically extracted and primary processing like relative coordinate calculation are done.

\begin{algorithm}[h]
	\caption{Script Record}
	\label{alg:record}
	\begin{algorithmic}[1]
		\Require Operation Sequence $OS$
		\Ensure Test Script $LITS$
		\State initiate $LITS$
		\For{each Operation $O$ $\in$ $OS$}
			\State initiate 7-dimension tuple $TS_o$
			\State $TS_o$ $\Leftarrow$ $DSN$
			\State $TS_o$ $\Leftarrow$ $O$
			\State $TS_o$ $\Leftarrow$ $SS_a$
			\State Crop the screenshot of operated widget $SS_w$
			\State $TS_o$ $\Leftarrow$ $SS_w$
			\State Extract the top-left and the bottom-right coordinate of the operated widget $C_w$
			\State $TS_o$ $\Leftarrow$ ($C_{wr}$ = $C_w$ / $DR$)
			\State Extract the coordinate of the operated point $C_o$
			\State $TS_o$ $\Leftarrow$ ($C_{or}$ = $C_o$ / $DR$)
			\State Extract the text on the widget $T$
			\State $TS_o$ $\Leftarrow$ $T$
			\State $LITS$ $\Leftarrow$ $TS_o$
		\EndFor 
		\State \Return $LITS$;
	\end{algorithmic}
\end{algorithm} 

\subsection{Script Replay}
For script replay, the proposed framework retrieves the script file from the database, and then orderly executes widget matching and single-step replay.

According to the formal expression in Algorithm \ref{alg:replay}, first, in the order of the operation sequences that obey the testing requirements, each step is replayed on the replaying devices. For each step, we extract the screenshot of the activity under test and the operated widget, and perform matching using both \image~and \layout~separately. \image~will output a set of possible widgets with \texttt{runImageMatching()} (Line 2 in Algorithm \ref{alg:replay}), and \layout~will output only one candidate widget with \texttt{runLayoutMatching()} (Line 3 in Algorithm \ref{alg:replay}). The nearest one in the possible widget set from \image~to the candidate widget from \layout~is considered as the candidate widget of \image (Line 4 to Line 17 in Algorithm \ref{alg:replay}). The 2 candidate widgets will be merged with parameter $\gamma$ to obtain a target widget.

Then, the coordinates of the target widget will be matched, and the operation information will be converted into executable commands on the replaying devices. A successful replay refers to the success of the operation on the right widgets and makes the app redirect to the preconceived activity.

 The proposed image-driven mobile app testing framework adopts two algorithms to match the operated widgets in the scripts and on the replaying devices: \textbf{\image} and \textbf{\layout}.

\textbf{\textit{Script Replay}}: For each step, the activity screenshot and widget information on the replaying device are extracted and compared with the information in the recorded $LITS$ script. Then the coordinate is acquired, and the corresponding operation is therefore done.

\begin{algorithm}[h]
	\caption{Script Replay}
	\label{alg:replay}
	\begin{algorithmic}[1]
		\Require Test Script $LITS$, Replaying Device $D_{replay}$
		\Ensure Test Result $TR$
		\State initiate target widget $W_{target}$
		\State $\Omega_{image}$ $\Leftarrow$ runImageMatching()
		\State $W_{target\_layout}$ $\Leftarrow$ runLayoutMatching()
		\If{$\Omega_{image}$.size() $==$ 1}
			\State $W_{target\_image}$ $\Leftarrow$ $\Omega_{image}$.get(0)
		\Else{
			\State $d_{min}$ $=$ $+\infty$
			\For{each widget $W_{image}$ $\in$ $\Omega_{image}$}
			\State $d_{widget}$ $\Leftarrow$ distance($W_{image}$, $W_{target\_layout}$)
			\If{$d_{widget}$ $<$ $d_{min}$}
				\State $d_{min}$ $\Leftarrow$ $d_{widget}$
				\State $W_{target\_image}$ $\Leftarrow$ $W_{image}$
			\Else
				\State continue
			\EndIf
		\EndFor}
		\EndIf
		\State set $\gamma$
		\State $TR$ $\Leftarrow$ operate($\gamma * W_{target\_image} + (1 - \gamma) * W_{target\_layout}$)
		\State \Return $TR$;
	\end{algorithmic}
\end{algorithm}

\begin{figure*}[!h]
	\centering
	\includegraphics[width=0.95\linewidth]{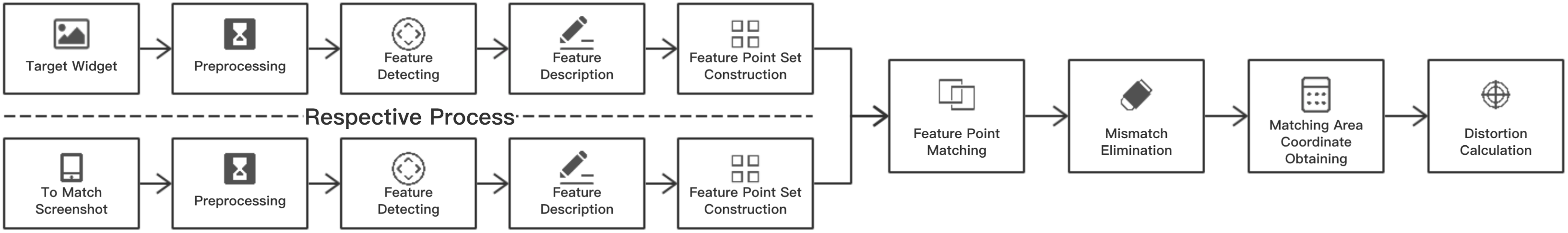}
	\caption{\image}
	\label{fig:ImageMatch}
\end{figure*}

\begin{figure*}[!h]
	\centering
	\includegraphics[width=0.75\linewidth]{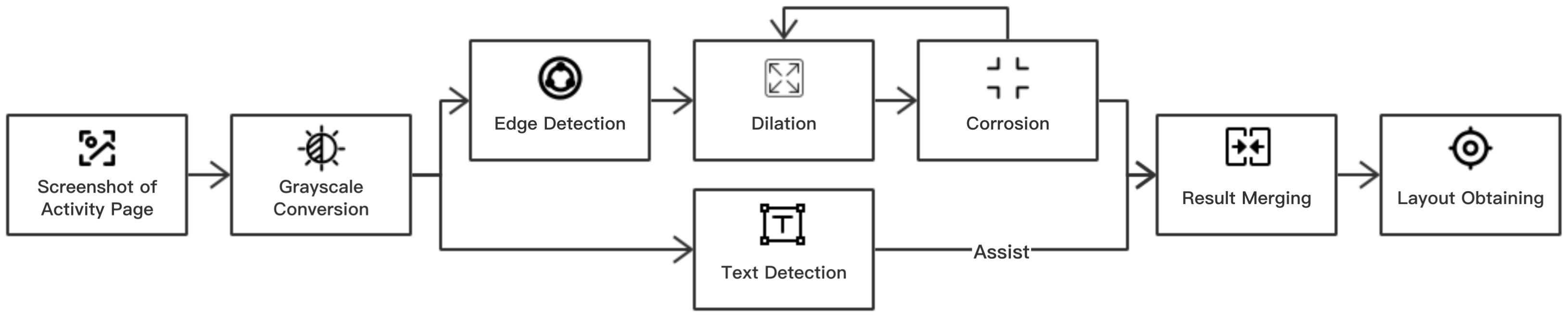}
	\caption{\layout}
	\label{fig:LayoutMatch}
\end{figure*}

\subsubsection{\textbf{\image.}}

\image~algorithm is used to match the screenshots of app widgets recorded in the scripts and the corresponding widgets on the replaying devices. It takes the target image and the image to match as input and outputs the coordinate value of the widgets. The algorithm includes five main processes: preprocessing, feature extraction, feature matching, mismatch elimination, and distortion calculation. Each process is described below:

\begin{itemize}
    \item \textbf{Preprocessing.} The input image of the widget is performed with grayscale processing since the color information of the image is not treated as a processing attribute. This process make a projection from a 3-channel image to an 1-channel, greatly improving the efficiency of subsequent processing and keeping the relative features of color changing and object contours \cite{lowe1999object}. The preprocessing enables a better effect in subsequent processing.

    \item \textbf{Feature Extraction.} This process includes detection of image features and construction of image feature descriptor set and feature point set. The target widget image and the activity image to match are processed separately, and two feature point sets (represented as $K_{target}$ and $K_{source}$), and two descriptor sets (represented as $D_{target}$ and $D_{source}$) are obtained.

    \item \textbf{Feature Matching.} We perform the feature matching and then estimate nearest neighbors. In the processing of two feature point sets, the nearest points found in $D_{target}$ and $D_{source}$ are considered as matching points. Then, the preliminary matching of the feature points is completed.

    \item \textbf{Mismatch Elimination.} Matching points may have mismatches, so it is necessary to eliminate such mismatches. We employ a ratio testing to address this problem. If the calculated value is smaller than a preset threshold $\delta$, the match is considered good. Otherwise, the match is considered as a mismatch and will be removed.

    \item \textbf{Distortion Calculation.} Since the target images may have distortions such as rotation and zoom, in order to obtain the position of the matching region more accurately, the homography matrix between the target widget image and the activity image to match is calculated. Finally, the perspective transformation of the target image is performed to obtain the coordinate position information.
\end{itemize}

\image~can complete the area matching of the widget screenshot to the image of the replaying device activity page. It can almost complete the processing from image input to coordinate output in a few milliseconds, which largely ensures the soundness in the replay process.

However, the limitation is that when the widget screenshots in the replaying devices change frequently or when there are many dynamic or similar widgets in the activity, the target widget can hardly be correctly positioned. Therefore, we will record all the suspected widget information and compare them with the result acquired from \layout, and finally calculate the probability for the suspect widgets to be operated.

\subsubsection{\textbf{\layout.}}

Because of the rapid refreshing of app contents, \image~would be out of effect because it relies heavily on the image features of app contents. In the UI testing tasks, the smoothness of app functionality, instead of the app contents, is the main concern. Therefore, a supplement of \image~positioning is necessary. Due to the similarity of app UI layouts among different platforms, we are considering further employing \layout~to improve the replay accuracy. Here we define the ``layout'' as the widget localization and the hierarchy relation among the widgets.

In the \layout, we first extract the widget contours based on the recorded activity screenshot stored in the scripts, and then divide the activity screenshot according to the widget contours, and acquire the relative position of the widgets on the activity. 
After obtaining all the widget contours, we will characterize the layout of the activity. First, we execute \textbf{\texttt{Group}} operation, which means a rough horizontal characterization to the activity. In this process, widgets wrapped in other widgets are omitted in this step, and we will get several groups of widgets. Then, we will divide each group into several lines by \textbf{\texttt{Line}} operation. In the \textbf{\texttt{Line}} operation, some widgets that wrapping other widgets will be segmented according to the contours of the wrapped widgets. Finally, in each line, we execute the \textbf{\texttt{Column}} operation to segment each line into several columns vertically. Therefore, each widget can be represented as a 3-tuple $(g, l, c)$, which means the group number, the line number and the column number. Also, the relative relationship among the widgets and the activity structure are also acquired according to the 3-tuple. The 3-tuple is platform-independent, and for replaying, the 3-tuples will be translated into 2-dimension coordinates according to the corresponding position in the \layout~results.

Our approach might generate some noise data. We also make efforts to eliminate such noise data. According to our survey on an open-sourced dataset \cite{yu2019crowdsourced}, we observe that the size of most widgets is more than 1\% of the screenshot size. Therefore, we discard the data whose size is smaller than the 1\% of the screenshot size.

In order to further improve the accuracy, we also extract text information on the widgets. On many occasions, some highly similar widgets with different texts are close to each other, making the matching hard to handle, so that the texts can assist the matching.

After the necessary information is collected, then starts the script replay process. 
During the replay, first, the same process is performed on the replaying devices, then we load the recorded information from the record device to match the information from the replaying devices.

With the \layout, we can easily solve the problem of dynamic widgets that the contents are rapidly refreshing. Take the news app as an example. While replaying the test script, the piece of news in the recorded script may have changed, and in the place of the news, there is a new piece of news. It is equivalent to click on the new piece of news. For \image, the 2 different pieces of new are definitely different, so the matching would fail. However, with the \layout, the framework will ignore the detailed content and pay attention to the widget position and the activity layout.

\section{Tool Implementation}
\label{sec:tool}

In order to devote the proposed image-driven mobile app testing framework into practice, we design and implement a tool, namely \toolname, which is short for \textbf{L}ayout and \textbf{I}mage \textbf{R}ecognition Driving Cross-Platform \textbf{A}utomated Mobile \textbf{T}esting. In this section, we illustrate the specific design and detailed algorithms and parameters. Based on \toolname, we also conduct an empirical experiment to evaluate the effectiveness of our proposed image-driven mobile app testing framework, which will be discussed in the next section.

\toolname is user-friendly. For script record, developers can select one specific device and operate on the webpage of the \toolname, and the operation will be automatically recorded and analyzed. For script replay, developers only need to invoke a recorded script, and then select the devices they want to execute the script replay. The following process is automatic.

\subsection{\image~Replay}

To replay with \image, we extract image features of activity screenshots and widget screenshots with SIFT algorithm \cite{lowe1999object}. The process can be seen in Fig. \ref{fig:ImageMatch}. The SIFT algorithm can effectively detect and describe local features in images, it is also a key technique adopted in state-of-the-art image-based record and replay tools. After extracting the image features and forming the feature point sets and descriptor sets (represented as $K_{target}$, $K_{source}$, $D_{target}$ and $D_{source}$), we employ FLANN library proposed by Muja et al. \cite{muja2009fast}. FLANN performs fast approximate nearest neighbor searches in high dimensional spaces. We perform FLANN algorithm on the extracted feature point sets from both the recorded activity screenshot under test and real-time activity on the replaying device. In the processing of two feature point sets, the KD-Tree index is used, and the KNN algorithm helps find the nearest points in $D_{target}$ and $D_{source}$ as matching points. Therefore, the preliminary matching of the feature points is executed. To eliminate mismatches of the matching points, we utilize a ratio testing method given by Lowe \cite{Lowe2004Distinctive}, which is calculated according to the formula $ratio=\frac{D_{min}}{D_{second\_min}}$. According to the practical experience, the threshold $\delta$ is set as 0.5 in our tool ($\delta$=0.5).

\subsection{\layout~Replay}

To solve the problems triggered by the drawbacks of the \image, we introduce the \layout. \layout~divides the activity screenshot into small areas according to the widget contours, and then uses the Canny algorithm to perform layout characterization. Then, we obtain the coordinate position information of the widgets on the recorded activity page of the recording device. Meanwhile, the same \layout with the Canny algorithm is performed on the activity screenshots of the replaying devices, and the target widget is positioned according to the extracted 3-tuple coordinate information. The main process can be seen in Fig. \ref{fig:LayoutMatch}. Since most apps have a similar layout for different versions on multiple platforms, and the test script model we propose is platform-independent, the cross-platform replay can be successfully realized.

We also refer to some other research work, like REMAUI \cite{Nguyen2016Reverse}, which is designed to generate code based on UI images. REMAUI uses Canny and OCR to characterize the image layout and combines the two algorithms to generate the page layout data structure. \toolname~encapsulates and improves the layout characterization approach REMAUI uses.

The Canny algorithm is elaborate on extracting edges, but if the detection for the contours is performed without extra processing, many tiny and redundant edge contours will be produced, which often have no significance in \layout. Instead, they can negatively affect the processing of contour data and layout characterization. Therefore, we expand the widget edges and connect the redundant contours to retain large, meaningful widgets, texts, etc. Contour detection is implemented by the ``\texttt{findContours()}'' function in the Canny algorithm, and the complete layout hierarchy is stored in the form of four vertex coordinates of the rectangular contour. Finally, we calculate the size of the extracted widgets, the ones which are smaller than 1\% of the activity size are discarded. We also introduce the OCR technology into \toolname to assist the matching.

However, the obtained widget set still has a lot of redundancy. Through the multiple empirical trials on different app activities, we conclude the following 2 rules to basically filter out redundancy, and to improve the effect.

\begin{itemize}
    \item Clear the contour with the length and width less than the pre-defined threshold in the contour. Practice from the analysis on large amount real-world apps shows that when the threshold is 2\% of the current device screen width, the contour element is not a functional widget.
    \item Clear the inner contour of the contour. (This rule is ignored when the contour is longer or wider than 60\% of the width of the corresponding device). When widgets occupy a small part of the device screens, the function of the inner widget is generally equivalent to the outer widget. Therefore, under such circumstances, such inner contour is meaningless.
\end{itemize}

After the layout characterization of the activity, each widget is assigned with a 3-tuple $(g, l, c)$ to represent its position.

Results both from \image~and \layout~are considered for final widget positioning. and we use a parameter $\gamma$ to calculate the final result, which can be seen in Line 19 of Algorithm \ref{alg:replay}.

\section{Empirical Evaluation}
\label{sec:verification}

Based on \toolname, we conduct an empirical experiment to evaluate the effectiveness of the proposed image-driven mobile app testing framework. In this experiment, we investigate to answer the following research questions:

\begin{itemize}
	\item \textbf{RQ1}: How effective can \toolname replay test scripts?
 	\item \textbf{RQ2}: How much can \toolname outperform the state-of-the-art image-based record and replay tools?
 	\item \textbf{RQ3}: Why does \toolname fail to replay in some cases?
\end{itemize}

\subsection{Experiment Setup}

To evaluate the effectiveness of the proposed image-driven mobile app testing framework and the tool \toolname, we define a matrix \textbf{\textit{Replay Accuracy}} to evaluate the effectiveness.

\textbf{\textit{Replay Accuracy}}: The percentage of the successful replays account for total replays.

\begin{table}[]
	\centering
	\caption{Experiment Mobile Application}
	\label{tbl:app}
	\begin{tabular}{cc|cc}
	
	\multicolumn{2}{c|}{Open-Sourced Apps} & \multicolumn{2}{c}{Commercial Apps} \\ \hline
	App Name        & Category     & App Name    & Category \\ \hline \hline
	AdGuard         & System       & Keep        & Sports   \\ 
	Jamendo         & Music        & Booking     & Shopping \\ 
	Kiwix           & Internet     & NBA App     & Sports   \\ 
	Linphone        & Phone \& SMS & Bing Search & Tool     \\ 
	Matomo Mobile 2 & Development  & Evernote    & Tool     \\ 
	Monkey          & Development  & McDonald    & Shopping \\ 
	openHAB         & Internet     & investing   & Finance  \\ 
	OsmAnd          & Map          & Taobao      & Shopping \\ 
	VLC             & Music        & QQ Music    & Music    \\ 
	WikiPedia       & Internet     & KFC         & Shopping \\ 
                    &              & Kindle      & Tool     \\
	\end{tabular}
\end{table}

\begin{table*}[]
	\centering
	\caption{Android Device in the Experiment}
	\label{tbl:device_a}
	\begin{tabular}{|c|c|c|c|c|c|c|c|c|c|}
	\hline
	
	Device ID & Serial No. & Usage & Brand & Market Name & Model & SDK & \tabincell{c}{Android \\ Version} & Resolution \\ \hline
	
	
	D0 & WBUBB18923510113 
	& Record & Huawei Honor & Honor 8X Max & ARE-AL10 
	& 27 & 8.1.0 & 1080 $\times$ 2244 \\ \hline
	
	
	D1 & 2003161a 
	& Replay & Oppo & Oppo & PBET00 
	& 27 & 8.1.0 & 1080 $\times$ 2340 \\ \hline
	
	
	D2 & 63fa9ed5 
	& Replay & Xiaomi & Xiaomi & MI 8 
	& 28 & 9     & 1080 $\times$ 2248 \\ \hline
	
	
	D3 & 7SWK89SO4HY5NJT8 
	& Replay & Vivo & Vivo & V1901A 
	& 28 & 9     & 720 $\times$ 1544  \\ \hline
	
	
	D4 & CLB0218724002507 
	& Replay & Huawei & Huawei P20 & EML-AL00 
	& 28 & 9     & 1080 $\times$ 2244 \\ \hline
	
	
	D5 & ce0717179034e124027e 
	& Replay & Samsung & Samsung & SM-N9508 
	& 28 & 9     & 1080 $\times$ 2220 \\ \hline
	
	\end{tabular}
\end{table*}
\begin{table*}[]
	\centering
	\vspace{-0.3cm}
	\caption{iOS Device in the Experiment}
	\label{tbl:device_i}
	\begin{tabular}{|c|c|c|c|c|c|c|c|c|c|}
	\hline
	Device ID 
	& UDID 
	& Usage 
	& Market Name 
	& Model 
	& OS Version 
	& Resolution 
	& UIKit Size
	\\ \hline
	
	D6 
	& \tabincell{c}{a81e386cf822ce0edeba741d \\
					64b04a8ca7d272e4}
	& Replay 
	& iPhone 6 
	& MG482LL/A
	& iOS 12.4.4
	& 750 $\times$ 1334 
	& 375 $\times$ 667
	\\ \hline
	
	D7 
	& \tabincell{c}{ba149a8863cee87c7dec7ec2 \\
					c6e4620e3f0568be}
	& Replay 
	& iPhone 7 
	& MNGX3CH/A
	& iOS 12.0(16A366)
	& 750 $\times$ 1334 
	& 375 $\times$ 667
	\\ \hline
	
	\end{tabular}
\end{table*}

During the selection of the experimental subjects, we consider the compatibility on multiple platforms. With this concern, we totally select 21 mobile apps, which can be referred to in TABLE \ref{tbl:app}. The selected apps include commercial ones and open-sourced ones, and cover 10 different categories, including \textit{system}, \textit{music}, \textit{internet}, \textit{phone \& SMS}, \textit{development}, \textit{finance}, \textit{tool}, \textit{sports}, \textit{shopping}, and \textit{map}, which can show the generalization capability. The selection is according to the ranking of Google Play and apps are filtered by the criteria of multiple platform supporting.

We also select multiple experimental devices for this experiment, including Android platform and iOS platform. The device list can be referred to in TABLE \ref{tbl:device_a} and TABLE \ref{tbl:device_i}. Our experiment covers most mainstream mobile brands and models, including Samsung, Huawei, Apple, Oppo, Vivo and Xiaomi. And the devices are of different operating system versions and different screen resolutions.

\subsection{RQ1: Script Replay}

\begin{figure}[!h]
	\centering
	\vspace{-0.2cm}
	\includegraphics[width=0.9\linewidth]{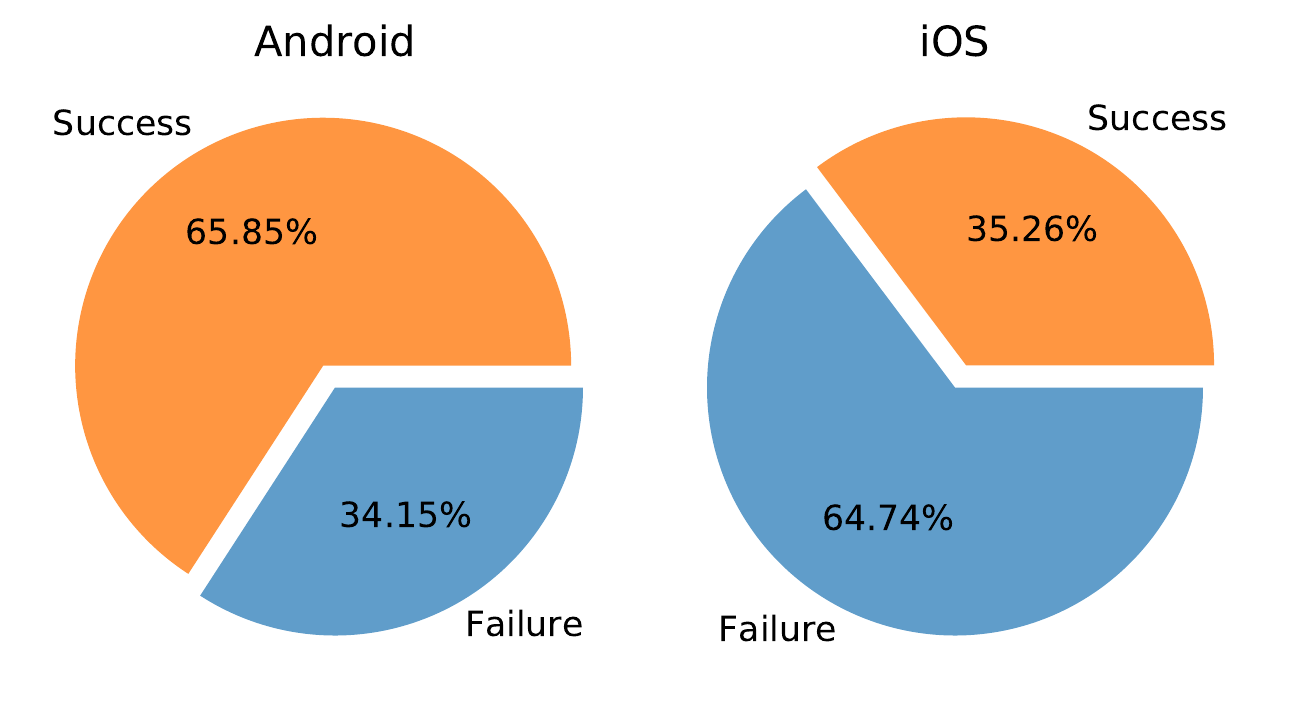}
	\caption{Replay Accuracy}
	\vspace{-0.2cm}
	\label{fig:RQ1}
\end{figure}

To answer RQ1, we recruit 4 senior software engineering majored students to organize 105 test scenarios on the experimental mobile apps, and record the test scenarios on an Android device labeled as D0. We require each test scenario includes 15 test operations, and each operation contains an app widget. Then, we simultaneously replay the recorded scripts on 5 Android devices and 2 iOS devices. As is shown in Fig. \ref{fig:RQ1}, the average replay accuracy of Android and iOS devices are around 65.85\% and 35.26\%\footnote{Kiwix and Jamendo are unusable on iOS devices due to the apps themselves.} respectively. The results show that the mobile app testing framework is promising. 

\begin{center}
\setlength{\fboxrule}{1pt}
	\fbox{
	\parbox{0.9\linewidth}{
	\textbf{\emph{The replay accuracy of Android device is around 65.85\%, and the replay accuracy of iOS device is around 35.26\%. Results show that the framework and \toolname to be promising.}}
	}
}\end{center}

\begin{figure*}[!h]
	\centering
	\includegraphics[width=\linewidth]{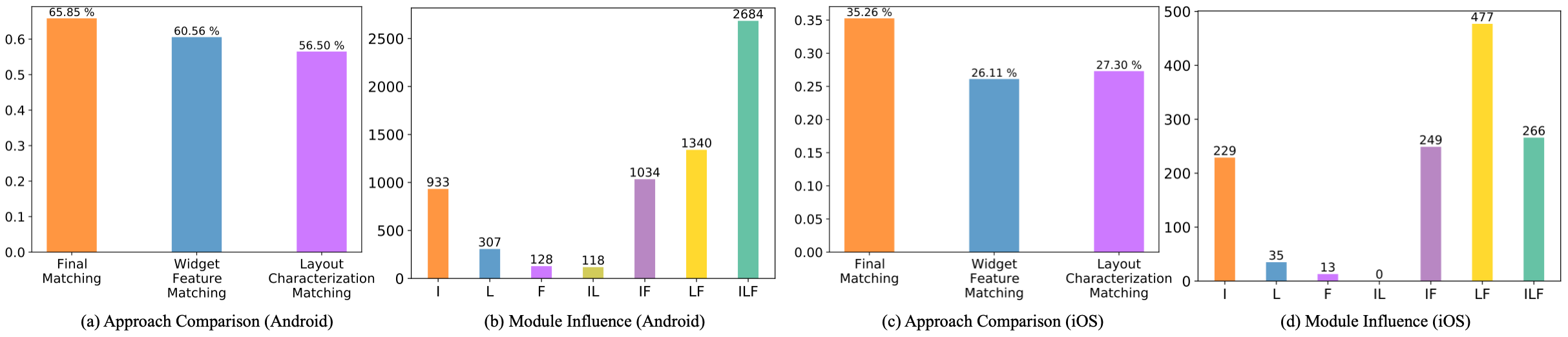}
	\caption{\textbf{Approach Comparison \& Algorithm Influence} (Labels containing ``\texttt{I}'' mean the \image~results are right; Labels containing ``\texttt{L}''mean the \layout~results are right and labels containing ``\texttt{F}'' mean the final results are right)}
	\label{fig:RQ2}
	\vspace{-0.3cm}
\end{figure*}

\subsection{RQ2: Baseline Comparison}

We further research on the comparison between \toolname and the state-of-the-art approaches. The final results are denoted as ``final results'', results from \image~are denoted as ``image results'' and results from \layout~are denoted as ``layout results''. Images results can be considered as the results of the state-of-the-art approaches because we obtain them with the same algorithms, and such tools are not available. Fig. \ref{fig:RQ2}(a) and Fig. \ref{fig:RQ2}(c) show the accuracy comparison among the final results, image results and layout results. Fig. \ref{fig:RQ2}(b) and Fig. \ref{fig:RQ2}(d) show the analysis of the influence among the final results, image results and layout results. There are 7 bars in subfigure (b) and (d). Labels containing ``\texttt{I}'' mean the image results are right; Labels containing ``\texttt{L}''mean the layout results are right and labels containing ``\texttt{F}'' mean the final results are right\footnote{We omit the situation when final results and results from 2 algorithms are all wrong}.

In the subfigure (a) and (c), \image~bar represents the results from the same algorithms adopted by state-of-the-art image-based record and replay approaches, such as Sikuli and Airtest. We can find that our framework has an improvement with \layout~over the state-of-the-art approaches by 8.74\% and 35\% on Android and iOS platform respectively. The improvements are especially obvious on cross-platform test script replay (from Android to iOS).

From the subfigure (b) and (d), we can find that with the combination of \image~and \layout, the replay accuracy is much higher than the two respective results. Especially for iOS, the increase is especially apparent. Moreover, compared with \image, the \layout~ also achieves a higher replay accuracy.

Among the results where the final results are right, 51.75\% and 26.47\% of successful replays (on Android and iOS) are due to the success from both algorithms; 19.94\% and 24.78\% of successful replays are due to the success from \image; 25.84\% and 47.46\% of successful replays are due to the success from \layout. However, even if 2 algorithms fail, there are 2.47\% and 1.29\% of final results to be successful. Among the data, we can find that \layout~can lead to much more final success than \image, which is 29.59\% on Android and 91.57\% on iOS. This phenomenon also proves that the introduction of \layout~to compromise the drawbacks of \image~alone achieves success, which is especially apparent in cross-platform replay.

\begin{center}
\setlength{\fboxrule}{1pt}
	\fbox{
	\parbox{0.9\linewidth}{
	\textbf{\emph{The introduction of \layout~greatly improves the replay accuracy compared with the same algorithms of the state-of-the-art image-based record and replay approaches. The improvements on Android and iOS platforms reach 8.74\% and 35\%. The improvement is especially apparent in cross-platform replay. Moreover, \layout~shows a more positive influence on the final replay accuracy.}}
	}
}\end{center}

\subsection{RQ3: Failure Analysis}

We review and analyze the failure cases one by one, and summarize some failure reasons.

In the \image, failures due to repeated highly similar widgets account for almost one-third of all failure cases, which is the most important reason for failure; secondly, the parsing failures in the recording phase result in the wrong screenshot of the target widgets. Such failures account for approximately 20\% of all failure cases. Some minor reasons have also led to individual failure cases, such as the missing of corresponding widgets on the replay device, too few feature points extracted by the algorithm, making the algorithm output result set empty.

In the \layout, 63\% failure cases are caused by layout characterization errors. Subtle layout changes caused by changes in activity contents caused approximately 16.3\% of failures. In addition, about 12.1\% of the failures are due to changes in the device status bar. Therefore, there is much space for improvement in layout characterization, and our approach will perform much better if the layout characterization algorithms has improvement.

\begin{center}
\setlength{\fboxrule}{1pt}
	\fbox{
	\parbox{0.9\linewidth}{
	\textbf{\emph{Several factors lead to replay failures, including repeat of highly-similar widgets, wrong widget screenshots, missing widgets on the replaying device, layout characterization error, layout changes led by activity changes or differences on status bar.}}
	}
}\end{center}

\subsection{Threats to Validity}

\textbf{The devices we use are limited}, we totally use 6 Android devices and 2 iOS device to complete the experiments. However, the mobile brand, model, and Android version have thousands of different types, thus the limitation of the device cluster can lead to a threat. However, the mobile devices we select are all the most popular ones on the current market, which accounts for a large percentage of the mainstream mobile device market.

\textbf{The representativeness of apps in our experiment is also a potential threat}. We select the popular apps that support both Android and iOS platforms. Even if we consider the diversity of the app category, we cannot cover all the categories. Also, some widely used apps that support only one single platform are also not considered. However, we think our proposed image-driven mobile app testing framework focuses on the UI of the mobile app, therefore, the different kinds of apps will not affect much. One important thing we have to claim that game apps that have no obvious layout are not applicable to the proposed framework.

\textbf{We recruit senior software engineering majored students to design the test scenarios and record test scripts in the experiment}. This may be a threat. However, Salman et al. propose that senior students are sufficient developer proxies in well controlled experiments \cite{salman2015students}.

\section{Related Work}
\label{sec:relatedWork}

\subsection{Code-Based Mobile Test Record and Replay} 

Traditional Android test script record tools such as Instrumentation \cite{instrumentation}, Robotium \cite{robotium}, UIAutomator \cite{uiautomator}, Espresso \cite{espresso} are some of the mainstream automated testing framework for Android platform. They encapsulate most operations and are easy to use. In iOS platform, KIF \cite{kif} and UIAutomation \cite{uiautomation} are the most widely used automated testing tools. However, the above tools have poor capabilities for cross-platform replay \cite{roy2014cross}, and the quality of test scripts depends largely on developers' capabilities.

Monkeyrunner \cite{monkeyrunner} is another automated testing tool on the Android platform. Monkeyrunner can replay test scripts on different devices simultaneously, greatly improving the test efficiency. However, users have to get familiar with the shell programming or python programming to write the test scripts, which is a demanding requirement and is unfriendly to users.

Appium \cite{appium} encapsulates different frameworks to support different platforms. Appium does not compile or adjust the app under test and can complete the automated test non-intrusively \cite{shah2014software}. However, the test scripts for different platforms cannot be generally used, so cross-platform replay is still hard to realize.

RERAN \cite{gomez2013reran} is a very early tool supporting Android test script record and replay. It captures events of low level with ADB by reading logs in ``\texttt{/dev/input/event*}'' files. Some similar tools like appetizer \cite{appetizer} and Mosaic \cite{halpern2015mosaic} utilize similar techniques. SARA, presented by Guo et al. \cite{guo2019sara} in 2019, further improves the Android application test script replay accuracy. However, such tools still cannot support cross-platform replay. Barista \cite{fazzini2017barista} proposed by Fazzini et al. is an approach that can help generate platform independent test cases while it start from the runtime state analysis.

The above tools are of high record and replay efficiency. However, they heavily rely on the platform frameworks and features. Even if some of them can support different platforms, like Android and iOS, developers still need to develop complete different test scripts for different platforms, and the developers need to know the different knowledge well, which is quite a high bar.

\subsection{GUI-Based Mobile Test Record and Replay} 

Behrang and Orso \cite{behrang2018automated} \cite{behrang2018test} propose AppTestMigrator, which allows for migrating test cases between apps with similar features using the similarity among GUI widgets. 

Sikuli, a tool presented by Yeh et al. \cite{yeh2009sikuli}, allows developers to use GUI element screenshots as a parameter. In the replay phase, Sikuli applies image retrieval algorithms to match widgets according to the screenshots in the scripts. Such ideas make Sikuli able to cross devices \cite{sun2018design}. However, the traditional computer vision algorithms it applies significantly limit the usability when used for cross-platform replay. UI elements become more dynamic and tend to have different scaling ratio, which is the drawback of pixel comparison method in the computer vision algorithm Sikuli uses.

Based on Sikuli, Airtest presented by NetEase is a cross-platform UI automated testing framework based on image recognition technology and \textit{Poco} widget recognition technology. Airtest improves the image recognition technology and adds a \textit{Poco} widget recognition technology in order to position the widgets by the \textit{XPath} or \textit{ID} values of the widgets. Airtest has higher accuracy in the replay phase. However, Airtest mainly focuses on the video games and has a relatively weaker support for a wider range of mobile apps. 

Moreover, some state-of-the-art approaches analyze the video information to record and replay test scripts within Android platform. Qian et al. \cite{ju2020roscript} propose a tool leveraging visual test scripts to express GUI actions and using a physical robot to drive automated test execution. Bernal-Cardenas et al. \cite{bernal2020translating} introduce V2S to translate video recordings of Android app usages into replayable scenarios.

The GUI based tools can alleviate the severity of the cross-platform problem. However, the problem is still not well solved. The depended image feature extraction and matching algorithms will meet quite much obstacles under the circumstances that app contents are constantly refreshed, leading to the constantly changing of image features.

\subsection{Widget Recognition Technology}

Chang et al. \cite{Chang2010GTU} present a new approach using computer vision technology for developers to automate their GUI testing tasks and execute all kinds of GUI behaviors.
Nguyen et al. \cite{Nguyen2016Reverse} firstly introduce an approach, namely REMAUI, to use input images to identify UI elements such as texts, images, and containers, using computer vision and optical character recognition (OCR) techniques. 

Moreover, Moran et al. \cite{moran2018machine} implement a tool on the basis of REMAUI, namely REDRAW, which can generate codes from UI images using the deep learning technology.
Additionally, Qin et al. \cite{qin2019testmig} present a tool, namely TestMig, for migrating test scripts from iOS to Android platform using widget hierarchy information and screenshots. 

Chen et al. \cite{chen2018ui} also present a neural network machine translator which combines recent advances in computer vision and machine translation, and translates UI images into GUI skeletons.
Chen et al. \cite{chen2020unblind} present an approach to automatically add labels to UI components using deep learning models.

Lowe \cite{lowe1999object} present an object recognition system, SIFT (scale-invariant feature transform), which uses a kind of news local image features. These features are invariant to image translation, scaling, and rotation, and partially invariant to illumination changes, affine or 3D projection.

Their work greatly enlightens us, and we absorb their ideas into widget recognition and test script record and replay in the image-driven mobile app testing framework.

\section{Conclusion}
\label{sec:conclusion}

Mobile apps often run on multiple platforms, so the limited capability of test scripts to work on multiple platforms can lead to repetitive developing work. The proposed image-driven mobile app testing framework solves such problems by the proposed platform-independent test script model and the \image~and \layout~algorithms, realizing the accurate positioning of the target widgets on different platforms. Our approach greatly simplifies the scripting work and makes it possible of ``one script record, multiple script replays'' on multiple platforms.

The experiment we conduct shows that the proposed image-driven mobile app testing framework achieves promising success in replaying mobile app test scripts on different platforms, and outperforms the state-of-the-art approaches much for cross-platform replay.

\section*{Acknowledgement}

This work is supported partially by National Natural Science Foundation of China (61932012, 61802171, 61772014), Fundamental Research Funds for the Central Universities (14380021), and National Undergraduate Training Program for Innovation and Entrepreneurship (202010284073Z).

\bibliographystyle{IEEEtran}
\bibliography{main}

\end{document}